\documentclass{PoS}
\usepackage{graphicx}% Include figure files
\usepackage[mathscr]{eucal}

\title{Absolute Measure of Local Chirality and the Chiral Polarization Scale of
       the QCD Vacuum 
       \thanks{A. Alexandru is supported in part by the U.S. Department of
               Energy under grant DE-FG02-95ER-40907. T. Draper is supported in
               part by the U.S. Department of Energy under grant
               DE-FG05-84ER40154.  The computational resources for this project
               were provided in part by the Center for Computational Sciences
               at the University of Kentucky, and in part by the George
               Washington University IMPACT initiative.  } }

\ShortTitle{Absolute Measure of Local Chirality}

\author{Andrei Alexandru$^{a},$\speaker{Terrence Draper}$^{b}$,
        Ivan Horv\'{a}th$^{b}$, and Thomas Streuer$^{c}$\\
        \llap{$^{a}$} George Washington University, 
                      Washington, DC 20052 \\
        \llap{$^{b}$} University of Kentucky, 
                      Lexington, KY 40506, USA \\
        \llap{$^{c}$} Institute for Theoretical Physics,
                      University of Regensburg,
                      93040, Regensburg, Germany 
        E-mail: \email{draper@pa.uky.edu}}

\abstract{The use of the absolute measure of local chirality is championed
since it has a uniform distribution for randomly reshuffled chiral components
so that any deviations from uniformity in the associated "X-distribution" are
directly attributable to QCD-induced dynamics. We observe a transition in the
qualitative behavior of this absolute X-distribution of low-lying eigenmodes
which, we propose, defines a chiral polarization scale of the QCD vacuum.}

\FullConference{The XXVIII International Symposium on Lattice Field Theory, Lattice2010\\
		June 14-19, 2010\\
		Villasimius, Italy}

\newcommand{\cP}{{\cal P}}                  % cal-P
\def\db{{\cP}_b}                            % dynamics in base coordinates
\def\eusm{\mathscr}
\def\Fg{{\eusm X}}                          % generic F
\def\Xg{X}                                  % generic X
\def\sgn{\mathop{\rm sgn}}                  % sign function

\begin{document}

\section{Introduction}

Important properties of the QCD vacuum can be studied indirectly by looking at
the characteristics of individual low-lying overlap fermion eigenmodes, which
provide a natural way to filter out UV fluctuations.  The space-time chiral
structure of the eigenmodes reflects the space-time topological structure of
the underlying gauge fields.  Properties of the
``$X$--distribution''\cite{Hor02}, the probability distribution of the local
chiral-orientation parameter, stirred a debate~\cite{Rabble} whether this
offered evidence in favor of, or against, models of the QCD vacuum.  We
revisited~\cite{Dra05, Ale10} the use of the local chiral-orientation parameter
and found that most of the qualitative features of the distribution were a
kinematical effect which was removed with the use of an improved and absolute
measure of local chirality.  Here we flesh out the construction of the
``absolute $X$--distribution'' and proceed to analyze the striking change in
its behavior as the energy scale of the eigenmodes is scanned~\cite{Ale10}.

\section{Local Chirality and the $X$--Distribution}

\begin{figure}[hb]
   \centering
   \vskip -0.10in
   \includegraphics[width=0.80\hsize,angle=0]{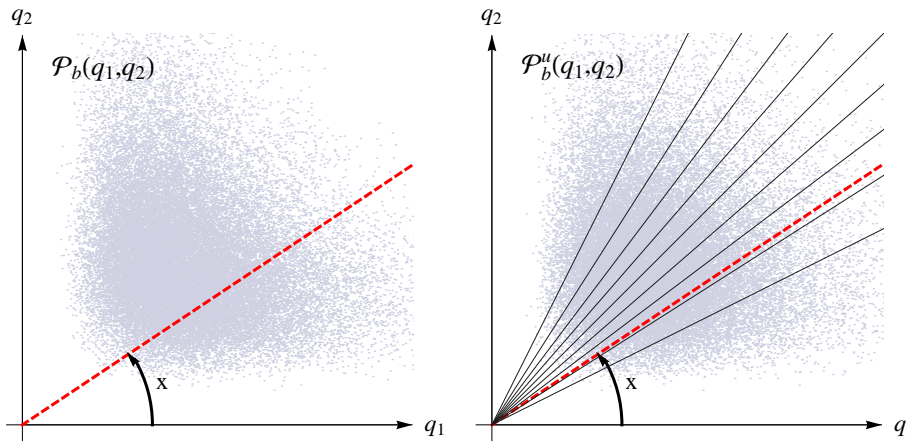} 
   \vskip -0.10in
   \caption{Left: Dynamics governing the polarization properties of two lowest
            modes for ensemble $E_1$ (defined in Table~1).  Right: The
            associated ``uncorrelated'' dynamics; solid gray lines separate the
            quadrant into 10 sectors, each containing 10\% of the population
            describing $\db^{u}$.}
   \vskip -0.10in
   \label{fig:dyn_E1}
\end{figure}

The local-chirality parameter measures the tendency of a low-lying Dirac
eigenmode, $\psi=\psi_L +\psi_R$ to be left or right handed.  Sampling the
space-time values of a particular set of eigenmodes, say at a particular energy
scale, for a set of configurations yields (a statistical approximation of) a
base probability distribution ${\cal P}_b(q_1,q_2)$ where $q_1\equiv|\psi_L|$
and $q_2\equiv|\psi_R|$.  A scatter plot of such a distribution for a
particular set of modes is shown in the left panel of Figure~\ref{fig:dyn_E1}.
We can elucidate the tendency for polarization by integrating over the radial
coordinate, $\rho$, and investigating the dependence on the polar angle,
$\varphi$.
It is convenient to symmetrize the angular variable $\varphi\in[0,\pi/2]$ by
defining the ``reference polarization coordinate'' $x \equiv
\frac{4}{\pi}\varphi-1$ so that $x \in [-1,+1]$ and such that a value of $x=+1$
($x=-1$) corresponds to purely right handed (left handed)
bispinors~\cite{Hor02}.  Thus, in the literature~\cite{Rabble}, histograms
which have a ``double-peak,'' that is two peaks near $x=\pm 1$ in a
distribution symmetric about $x=0$, have been interpreted (prematurely as we
will see) as revealing that the eigenmodes display chiral polarization.  The
amount of peaking can be enhanced with a different choice of ``polarization
function,'' that is, using functions of $x$, rather than $x$ itself.
These polarization functions can be classified as belonging to one of several
families, such as
\begin{equation}
    \Fg^C(x;\alpha) = \sgn(x) \; |x|^{\alpha}
    \quad ; \quad
    \Fg^R(t;\alpha) = \frac{t^\alpha - 1}{t^\alpha + 1}  
    \quad ; \quad
    \Fg^G(t;\alpha) = \frac{4}{\pi} \tan^{-1}(t^\alpha) - 1  
\end{equation}
where $\alpha > 0$ and $t\equiv \tan(\varphi)$.  $\Fg^C(x;1)=x=\Fg^G(t;1)$
correspond to the original chiral orientation parameter~\cite{Hor02}.
$\Fg^G(t;2)$ and $\Fg^R(t;2)$ are alternatives that appeared in the
literature~\cite{Rabble}.  All of these choices of polarization functions share
the same crucial features of the simplest choice of the reference polarization
function $\Fg(x)=x$, namely they are odd, monotonically-increasing functions.

\begin{figure}[hb]
  \centering
  \vskip -0.00in
  \includegraphics[width=0.46\hsize,angle=0]{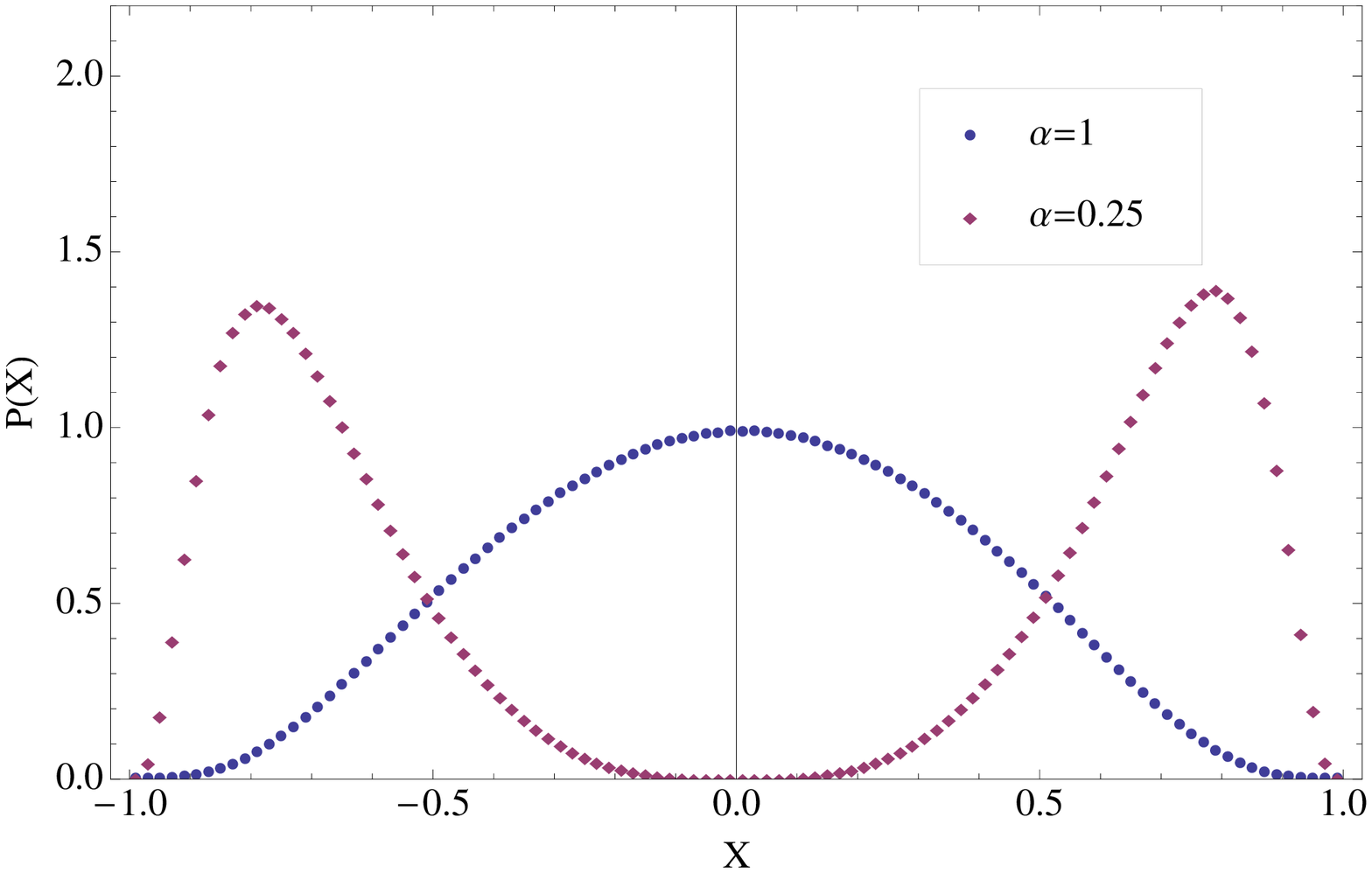}
  \includegraphics[width=0.46\hsize,angle=0]{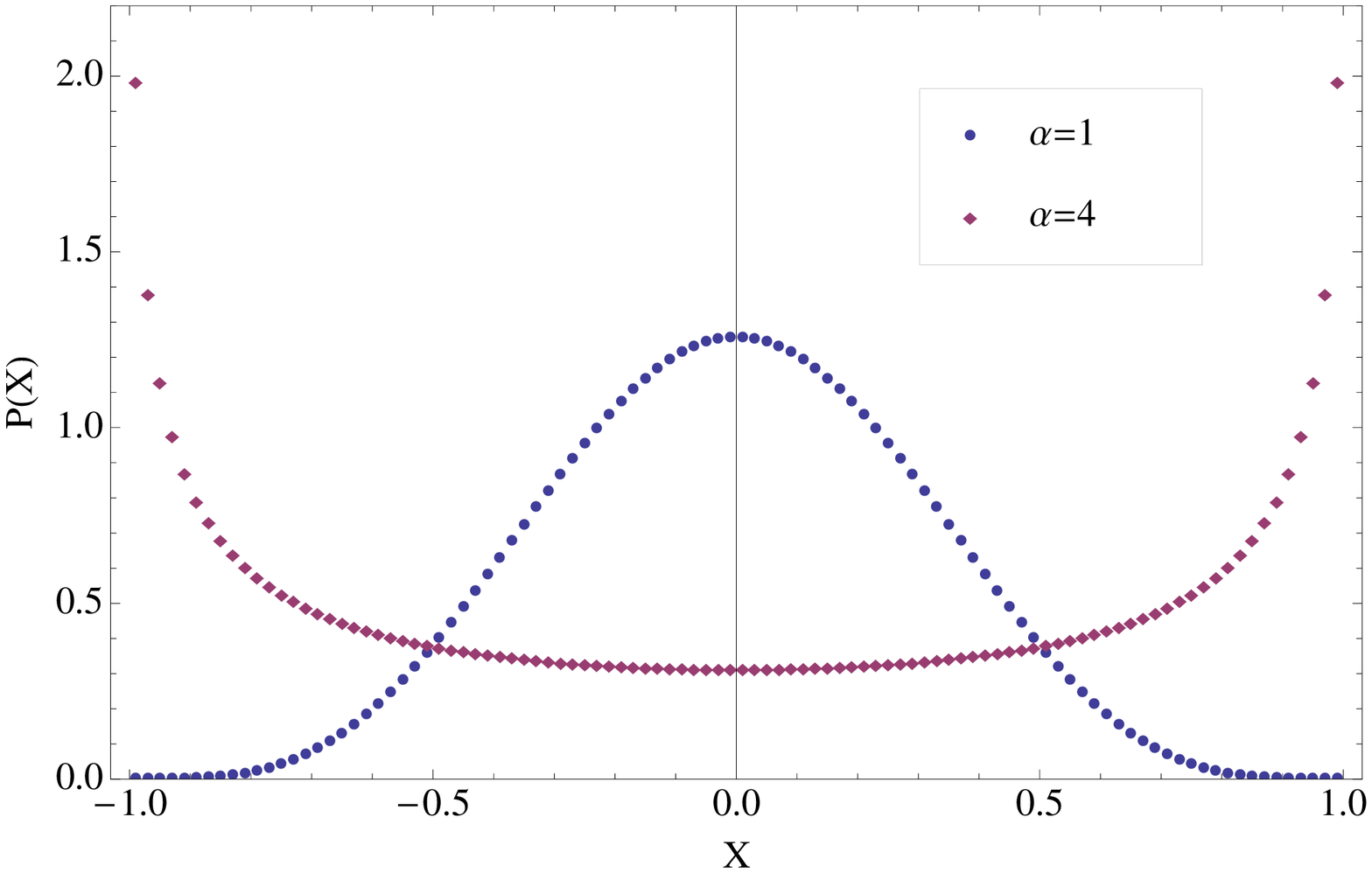}     
  \vskip -0.20in
  \caption{Possible $X$--distributions associated with fixed dynamics can have
           qualitatively different behavior. Here we select two polarization
           functions from family $\Fg^C$ (left), and two from family $\Fg^R$
           (right). In this and subsequent figures, the data have error bars
           but they may to be too small to be resolved.}
  \vskip -0.10in 
  \label{fig:arbitrary}
\end{figure} 

The marginal distribution
\begin{eqnarray}
    P(X) & = &  
      \int_0^\infty d q_1 \int_0^\infty d q_2 \,
%      {\cal P}_b(q_1,q_2) \; \delta\Bigl( X - \Fg(q_1,q_2) \Bigr)
      \db(q_1,q_2) \; \delta\Bigl( X - \Fg(q_1,q_2) \Bigr)
      \label{eqn:marginal}
\end{eqnarray}
where $X$ is a generic independent variable parametrizing the range of
polarization functions, i.e. $X\in [-1,+1]$, is called the
``$X$--distribution''~\cite{Hor02}.  It depends not only on the dynamical base
probability ${\cal P}_b(q_1,q_2)$ but also on the polarization function
$\Fg(x)$ used to measure it.  As Fig.~\ref{fig:arbitrary} indicates, the
qualitative features of the $X$--distributions depend very strongly on the
choice of the polarization function, and one might be led into declaring,
perhaps falsely, that the distribution is highly polarized or highly
unpolarized depending on ones arbitrary choice.

\begin{figure}[ht]
  \centering
  \vskip -0.00in
  \includegraphics[width=0.46\hsize,angle=0]{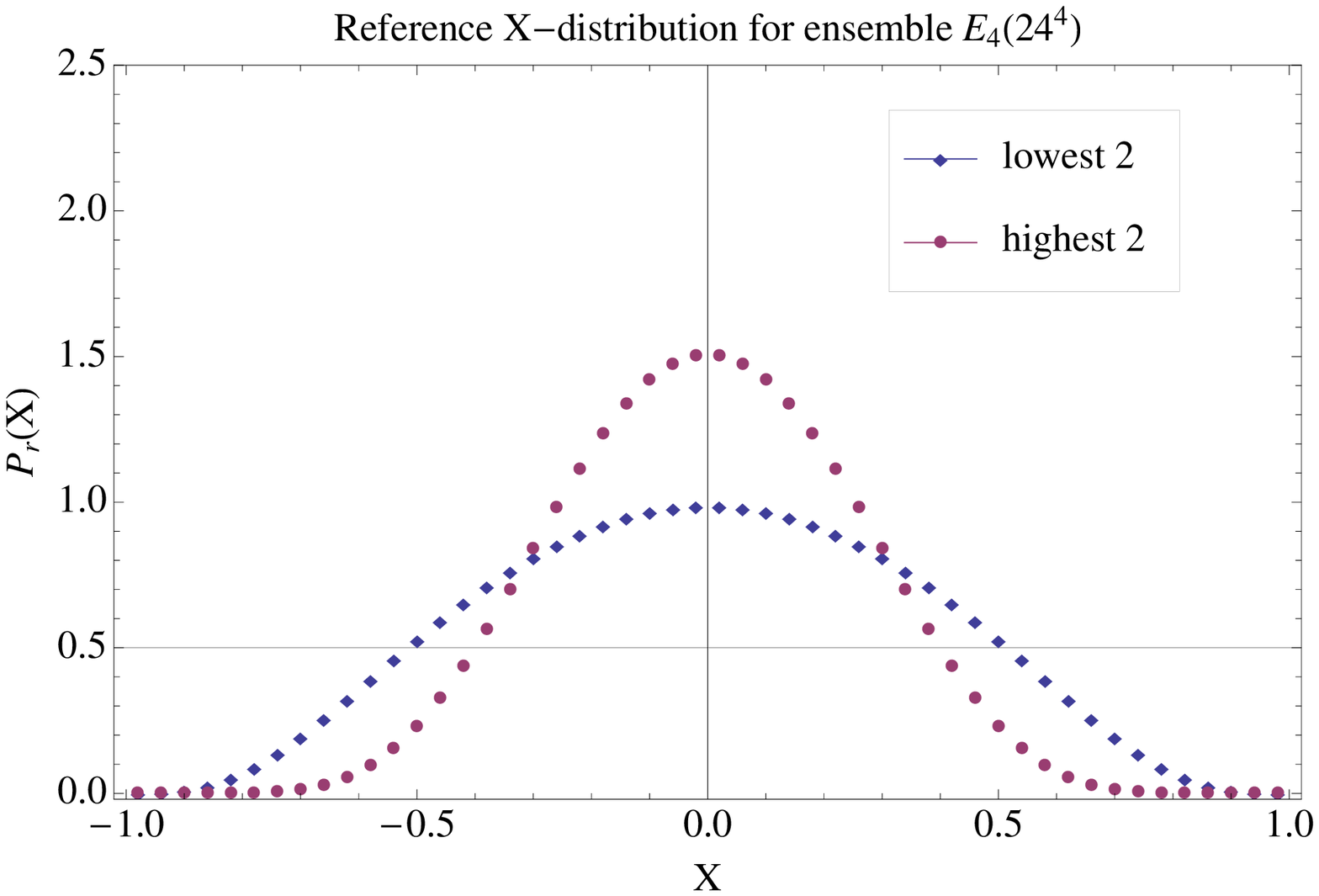}
  \includegraphics[width=0.46\hsize,angle=0]{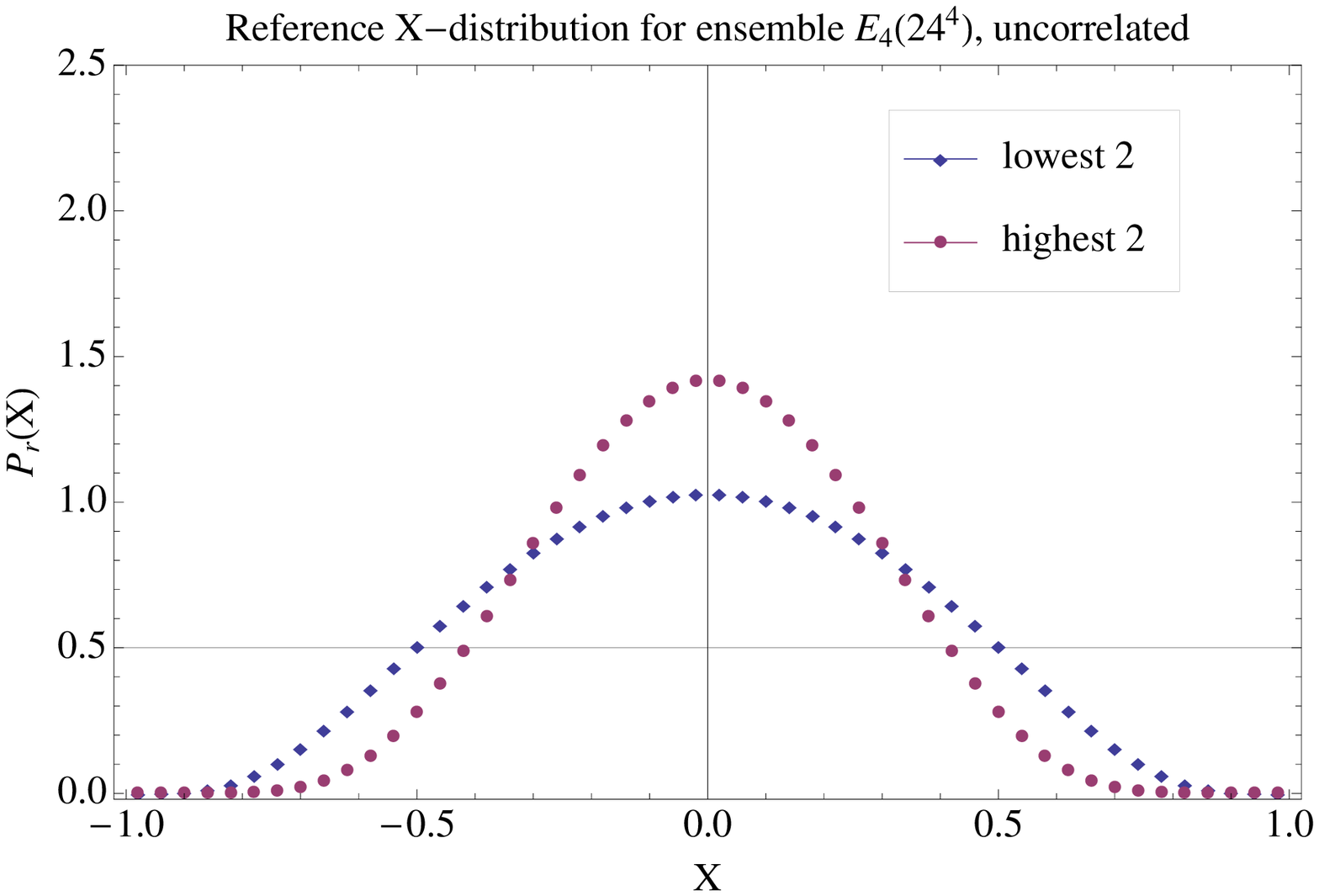}     
  \vskip -0.20in
   \caption{The reference $X$--distribution for ``lowest'' and ``highest''
            modes in ensemble $E_4$ (defined in Table~1) is shown in the left
            panel. The associated distribution with statistically independent
            left--right components (``uncorrelated'') is shown in the right
            panel.}
  \vskip -0.10in 
  \label{fig:pr-low-high}
\end{figure}

Indeed, most of the qualitative appearance of an $X$--distribution is a result
of "kinematics" (or "phase space") and is not due to QCD correlations at
all~\cite{Dra05}. To demonstrate this, we compute the $X$--distribution arising
from $\psi_L(y)$ and $\psi_R(y)$, where $y$ is the spacetime coordinate, and
then randomly reshuffle the fields $\psi_{Lran}(y)\equiv\psi_L(y_{ran})$, where
$y_{ran}$ is a random permutation, using an independent random reshuffle for
$\psi_R(y)$. Any QCD-dynamically-induced correlation between $\psi_L(y)$ and
$\psi_R(y)$ at site $y$ is destroyed by this procedure.  Nevertheless, the
correlated and uncorrelated scatter plots of the base probability distributions
are very similar as is seen in Fig.~\ref{fig:dyn_E1}. Accordingly, when we
re-compute the $X$--distribution, the randomized uncorrelated $X$--distribution
looks very similar to the original as seen in Fig.~\ref{fig:pr-low-high}.

\section{The Absolute $X$--Distribution}

To expose the true correlation we need to "subtract" the phase space
background, or more precisely measure the (QCD-dynamically-generated)
correlated distribution {\it relative\/} to the (randomized) uncorrelated one.

\begin{figure}[b]
  \centering
  \vskip -0.20in
  \includegraphics[width=0.70\hsize,angle=0]{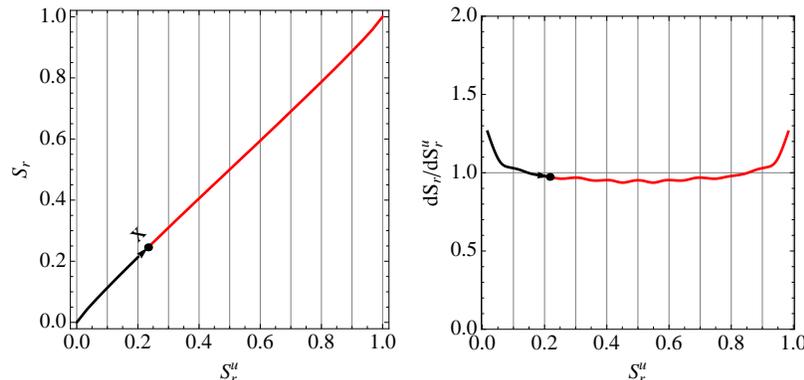} 
  \vskip -0.30in
  \caption{Steps in the construction of absolute $X$--distribution for data
           shown in Fig.~1. See the discussion in the text.  }
  \vskip -0.00in
  \label{fig:sr_vs_sru}
\end{figure}

The following procedure makes a differential comparison of polarization in the
dynamics ${\cal P}_b(q_1,q_2)$ to that of uncorrelated probability distribution
${\cal P}_b^u(q_1,q_2)\equiv p(q_1) p(q_2)$, where
$p(q_1)=\int_{-1}^{+1}dq_2{\cal P}_b(q_1,q_2)$ is a marginal distribution: A
ray that passes through the origin is specified by its reference polarization
coordinate $x$, as shown in Fig.~\ref{fig:dyn_E1}. Determine the fraction of
population contained between the $q_{1}$-axis and this ray, separately for
${\cal P}_b(q_1,q_2)$ and for ${\cal P}_b^u(q_1,q_2)$. These fractions are the
cumulative probability functions $S_{r}(x)=\int_{-1}^x dx' P(x')$ and
$S_{r}^{u}(x) = \int_{-1}^x dx' P_u(x')$.  Eliminating $x$, every ray is
represented by a point in the cumulative distribution plane $(S_r^u, S_r)$ with
the set of all such points forming a curve starting at $(0,0)$ and ending at
$(1,1)$. If the the two dynamics in question have identical polarizations the
graph will be a straight line.  On the left side of Fig.~\ref{fig:sr_vs_sru} we
plot the result of this construction for data displayed in
Fig.~\ref{fig:dyn_E1}. The correlated and uncorrelated dynamics have very
similar polarizations since the graph is close to being linear.  To obtain a
differential comparison, and to better see the differences, we compute the
slope of the cumulative polarization graph and show it on the right side of
Fig.~\ref{fig:sr_vs_sru}. As one can clearly see now, the correlated dynamics
exhibits a small excess of polarization with respect to the uncorrelated one
near the extremal values. This is a plot of the ``absolute $X$--distribution''
\begin{equation}
  P_A = \frac{1}{2}\frac{dS_{r}}{dS_{r}^{u}}
\end{equation}
except for the additional rescaling $S_r^u \rightarrow X \equiv 2 S_{r}^{u}-1$
required since $X\in[-1,+1]$ while the uncorrelated cumulative probability
$S_{r}^{u}\in[0,+1]$.  This construction is manifestly independent of the
choice for polarization coordinate, with different possibilities corresponding
to different parametric representations of the same curve in the
$(S_{r}^{u},S_{r})$ plane.

\begin{figure}[ht]
  \centering
  \vskip -0.00in
  \includegraphics[width=0.46\hsize,angle=0]{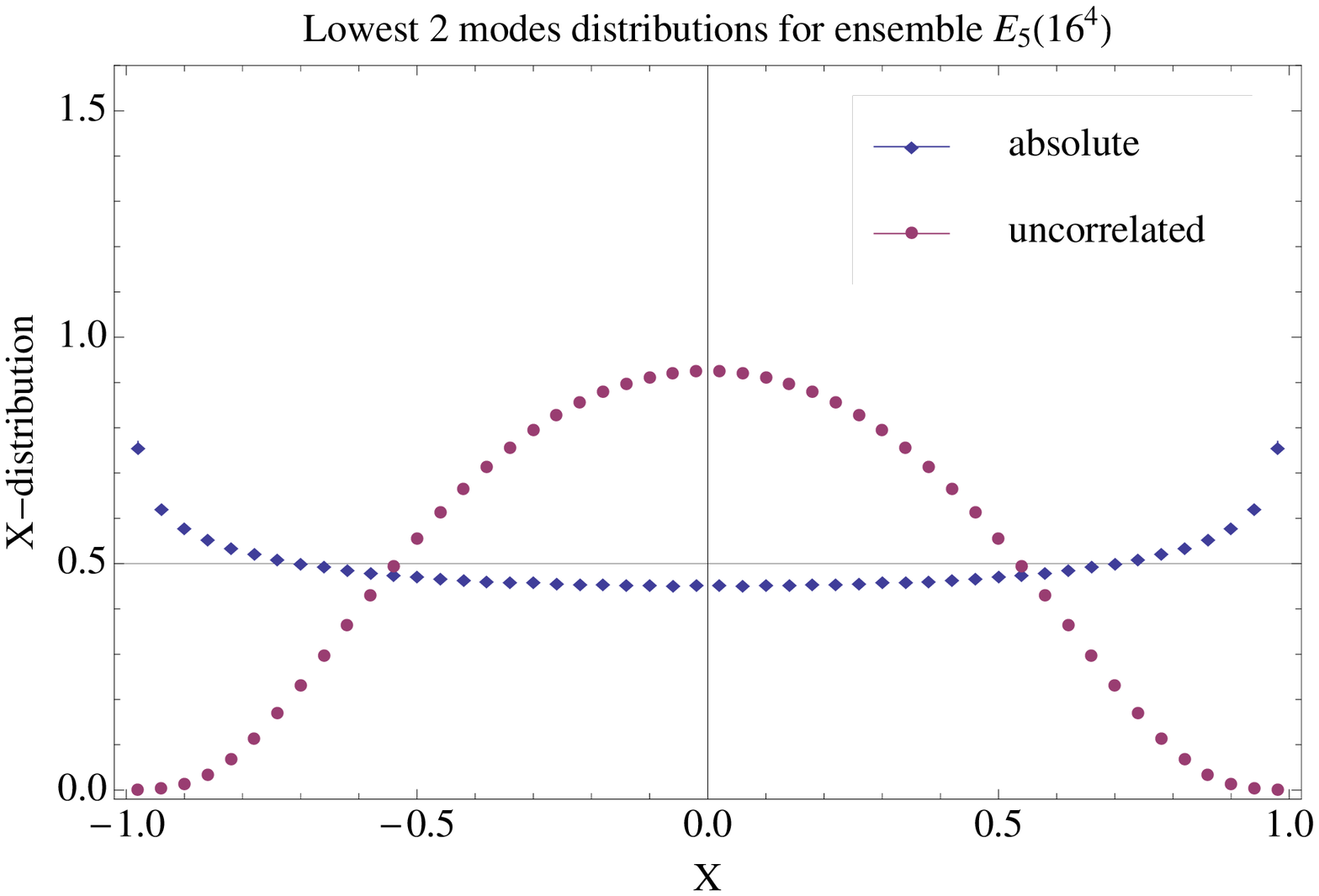}
  \includegraphics[width=0.46\hsize,angle=0]{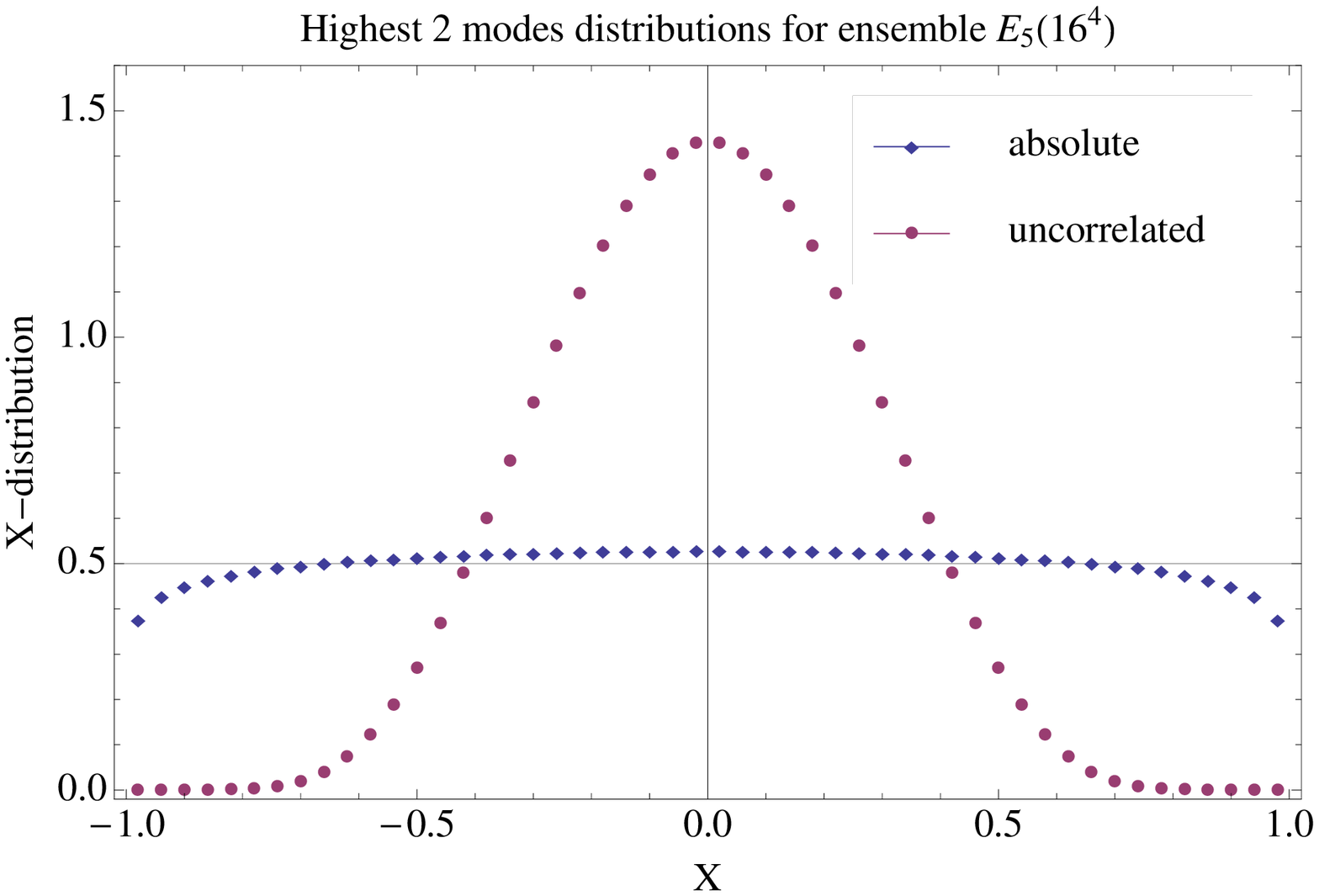}     
  \vskip -0.10in
  \caption{Absolute (dynamical) and uncorrelated (kinematical)
           $X$--distributions for gauge ensemble $E_5$ (defined in Table~1)
           using the reference polarization coordinate for the latter.  Results
           for lowest two non--zero pairs are shown on the left while the
           highest two pairs are shown on the right.}
  \vskip -0.20in 
  \label{fig:E5}
\end{figure}

Fig.~\ref{fig:E5} shows the absolute $X$--distribution $P_A(X)$ for the lowest
two non-zero pairs (left panel) and the highest two (of those measured) from
ensemble $E_5$ (defined in Table~\ref{Table:ensembles}).  In comparison is
shown the uncorrelated $X$--distribution, $P_r^u(x)$, constructed using the
reference polarization coordinate $x$. The shape of $P_A(X)$ is determined by
the correlation induced by QCD dynamics, while the shape of $P_r^u(x)$ is
determined completely by the kinematics.  From the pair, $P_A(X)$ and
$P_r^u(x)$, one could reconstruct the correlated (and uncorrelated)
$X$--distributions for {\it any\/} choice of polarization function $\Fg(x)$,
such as those chosen in~\cite{Rabble}.

\section{Convexity or Concavity of Absolute $X$--Distributions}

Now we look at the $X$--distributions for the lowest and highest modes (of
those measured low-lying modes) for our finest ($24^4$) ensemble ($E_4$).  We
have a total of four ensembles on lattices at various lattice spacings, with
fixed volume, and a fifth to check for finite-size effects.  These are listed
in Table~\ref{Table:ensembles}.

\newcommand\fm{\mathop{\rm fm}}
\begin{table}[ht]
  \centering
  \begin{tabular}{@{} ccccccccc @{}} % Column formatting, @{} suppresses leading/trailing space
    \hline
    Ensemble & Size & $N_{\rm config}$ & Volume & Lattice Spacing & 
    $\Lambda_{\rm LOW}^{\rm MAX}$ & $\Lambda_{\rm LOW}^{\rm AVE}$ & 
    $\Lambda_{\rm HIGH}^{\rm MIN}$ & $\Lambda_{\rm HIGH}^{\rm AVE}$\\
    \hline
    $E_{1}$ &  $8^{4}$ & 100 & $(1.32\fm)^{4}$ & $0.165 \fm$ & 449 & 226 & 1956 & 1980\\
    $E_{2}$ & $12^{4}$ &  97 & $(1.32\fm)^{4}$ & $0.110 \fm$ & 407 & 169 & 1711 & 1735\\
    $E_{3}$ & $16^{4}$ &  99 & $(1.32\fm)^{4}$ & $0.0825\fm$ & 304 & 142 & 1513 & 1553\\
    $E_{4}$ & $24^{4}$ &  96 & $(1.32\fm)^{4}$ & $0.055 \fm$ & 344 & 136 & 1338 & 1366\\
    $E_{5}$ & $16^{4}$ &  99 & $(1.76\fm)^{4}$ & $0.110 \fm$ & 162 &  58 & 1087 & 1123\\
    \hline
  \end{tabular}
  \caption{The summary of five ensembles used in overlap eigenmode
           calculations. The right side of the table describes some properties
           of the spectra (in MeV) with $\Lambda_{\rm LOW}^{\rm AVE}$ denoting
           the average magnitude of lowest near--zero eigenvalue over the
           ensemble and $\Lambda_{\rm HIGH}^{\rm AVE}$ denoting the same for
           highest eigenvalue. $\Lambda_{\rm LOW}^{\rm MAX}$ is the magnitude
           of the maximal lowest eigenvalue, and $\Lambda_{\rm HIGH}^{\rm MIN}$
           the magnitude of the minimal highest eigenvalue.}
  \label{Table:ensembles}
\end{table}

\begin{figure}[hbt]
  \centering
  \vskip -0.00in
  \includegraphics[width=0.46\hsize]{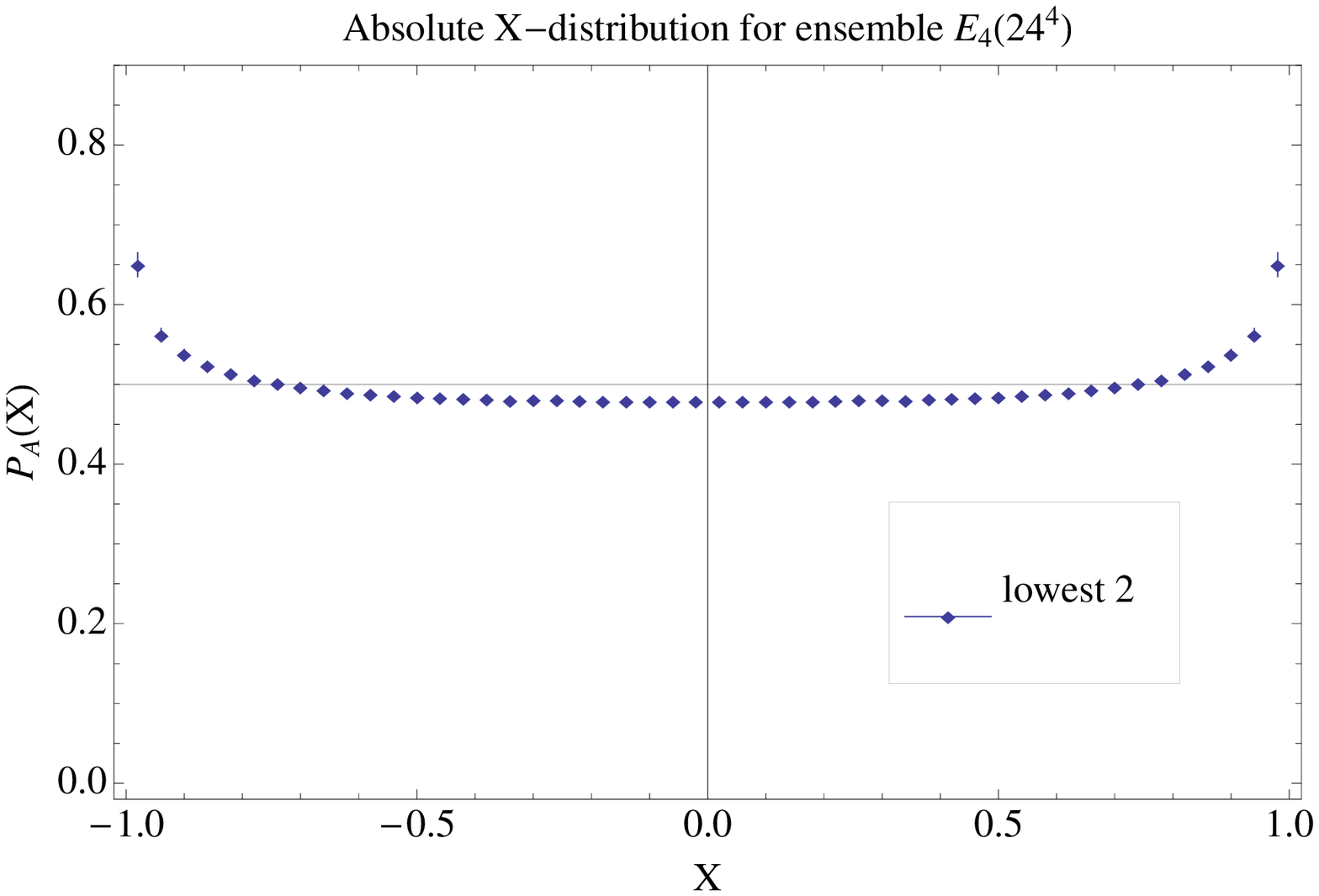}
  \includegraphics[width=0.46\hsize]{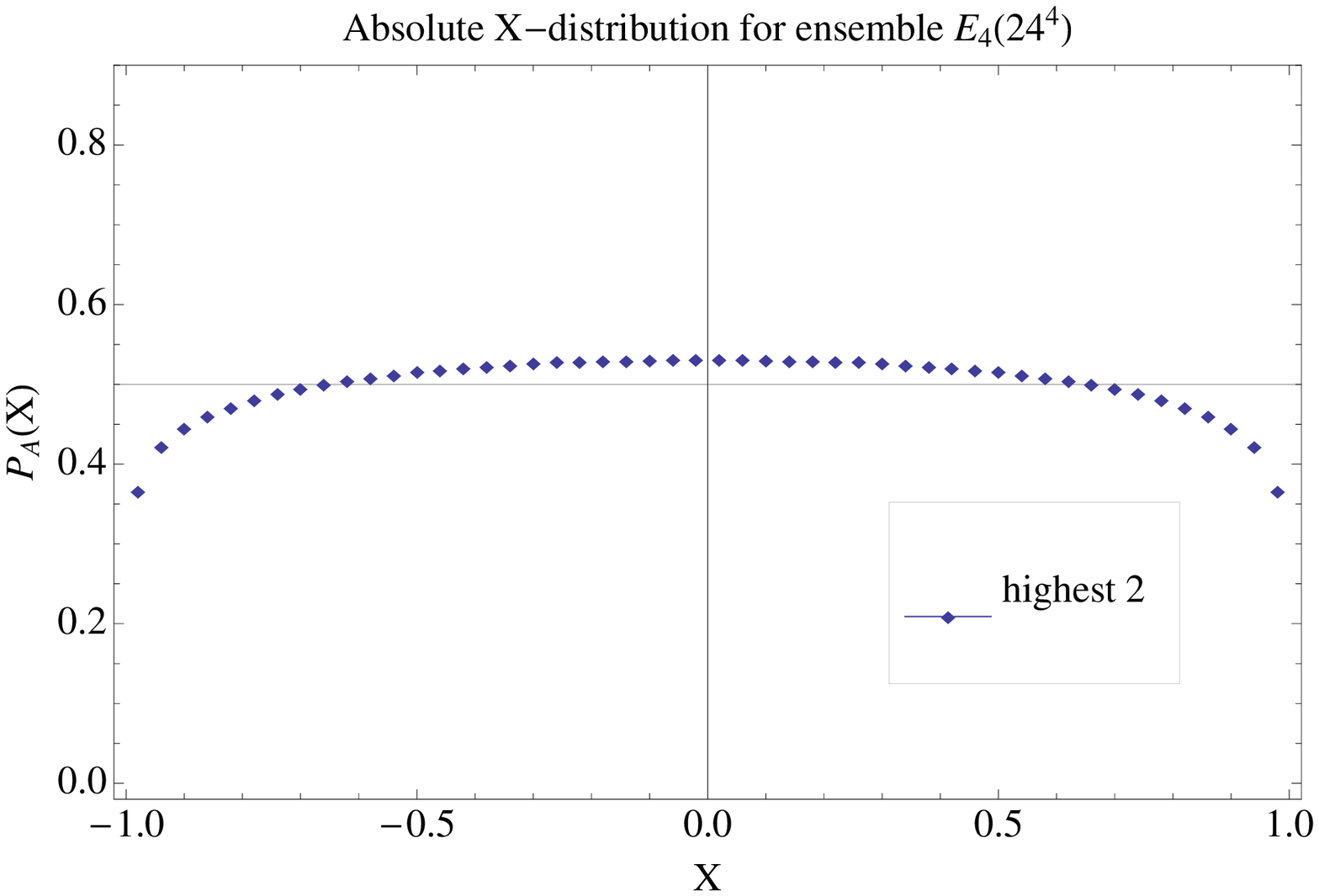}
  \vskip -0.20in
  \caption{A plot of the absolute $X$--distribution for the ``lowest'' (left
           panel) and ``highest'' (right panel) modes for our finest
           ensemble. }
  \vskip -0.10in
  \label{fig:XDist}
\end{figure}

Figure~\ref{fig:pr-low-high} (left panel) shows the (reference)
$X$--distribution for the lowest pair of eigenmodes and for the highest pair
(of the 55 measured).  ``Subtracting'' from this the corresponding
$X$--distribution for randomized data leaves the absolute $X$--distribution,
$P_A(X)$, shown Fig.~\ref{fig:XDist}.  We see that much of the peaking in the
reference $X$--distribution has been removed, by the comparison to the
uncorrelated version.  This indicates that most of the peaking was kinematical
rather than dynamical in the original reference distribution and that to draw
any conclusions from this distribution would be misleading at best.  On the
other hand, for the absolute $X$--distribution, any deviations from uniformity
are attributable to correlations induced by QCD dynamics. There is a relatively
small but clear effect exhibited by the lowest modes which have a convex shape
and the higher modes which have a concave $X$--distribution.  That is to say,
the lowest modes exhibit an enhanced chiral polarization while the higher modes
exhibit a suppression, which we regard as a novel and tantalizing discovery.

\section{Chiral Polarization Scale}

\begin{figure}[t]
\begin{center}
  \centering
  \vskip -0.00in
  \includegraphics[width=0.75\hsize,angle=0]{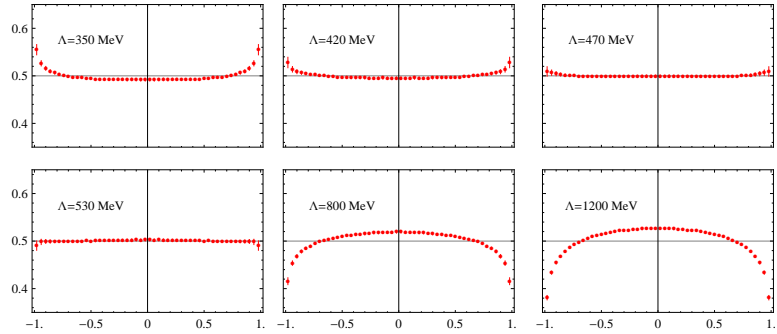}
  \vskip -0.25in
  \caption{Absolute $\Xg$--distributions for gauge ensemble $E_4$ with changing
           scale $\Lambda$ of the modes. Note the transition from convex to
           concave (positive to negative correlation) between 470 MeV and 530
           MeV.}
  \vskip -0.20in 
  \label{fig:pa-E4-alllam}
\end{center}
\end{figure} 

From Figs.~\ref{fig:XDist} and~\ref{fig:pa-E4-alllam} we see that the shape of
the absolute $X$--distribution curve is qualitatively different for the lowest
several modes (convex) versus higher modes (concave).  We seek the scale
$\Lambda_T$ at which the transition from convex to concave occurs.  To do this
we first define, as a figure of merit, a dynamical correlation coefficient in
terms of the first moment of the absolute $X$--distribution.
\[
C_A(X)=-1+2\int^{1}_{-1}dX |X| P_A(X)
\]
for which $C_A>0$ for enhanced polarization, $C_A<0$ for suppressed
polarization, and $C_A=0$ for neutrality as for the case of statistical
independence.

\begin{figure}[ht]
  \centering
  \vskip -0.00in
  \includegraphics[width=0.46\hsize]{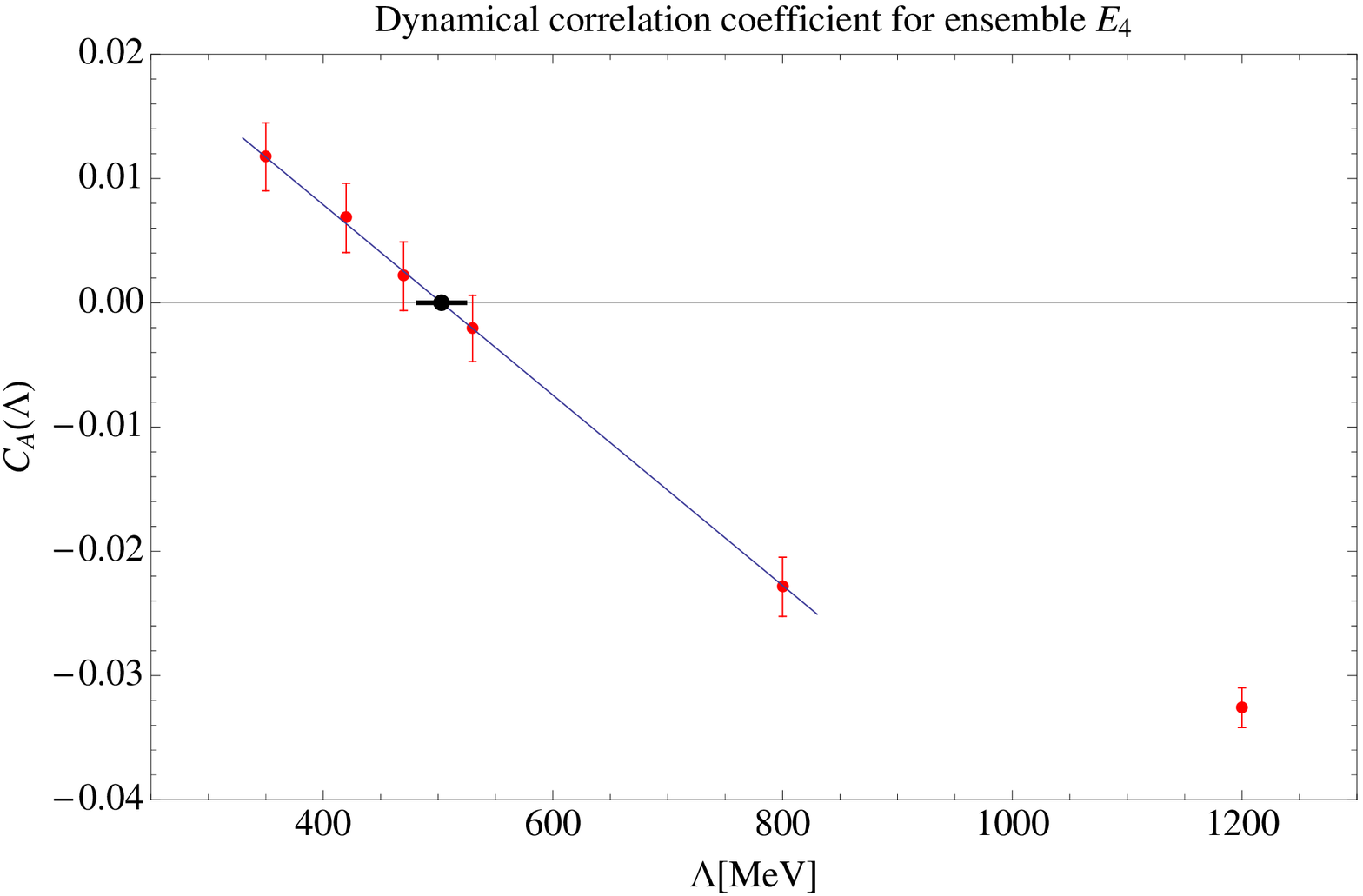}
  \includegraphics[width=0.46\hsize]{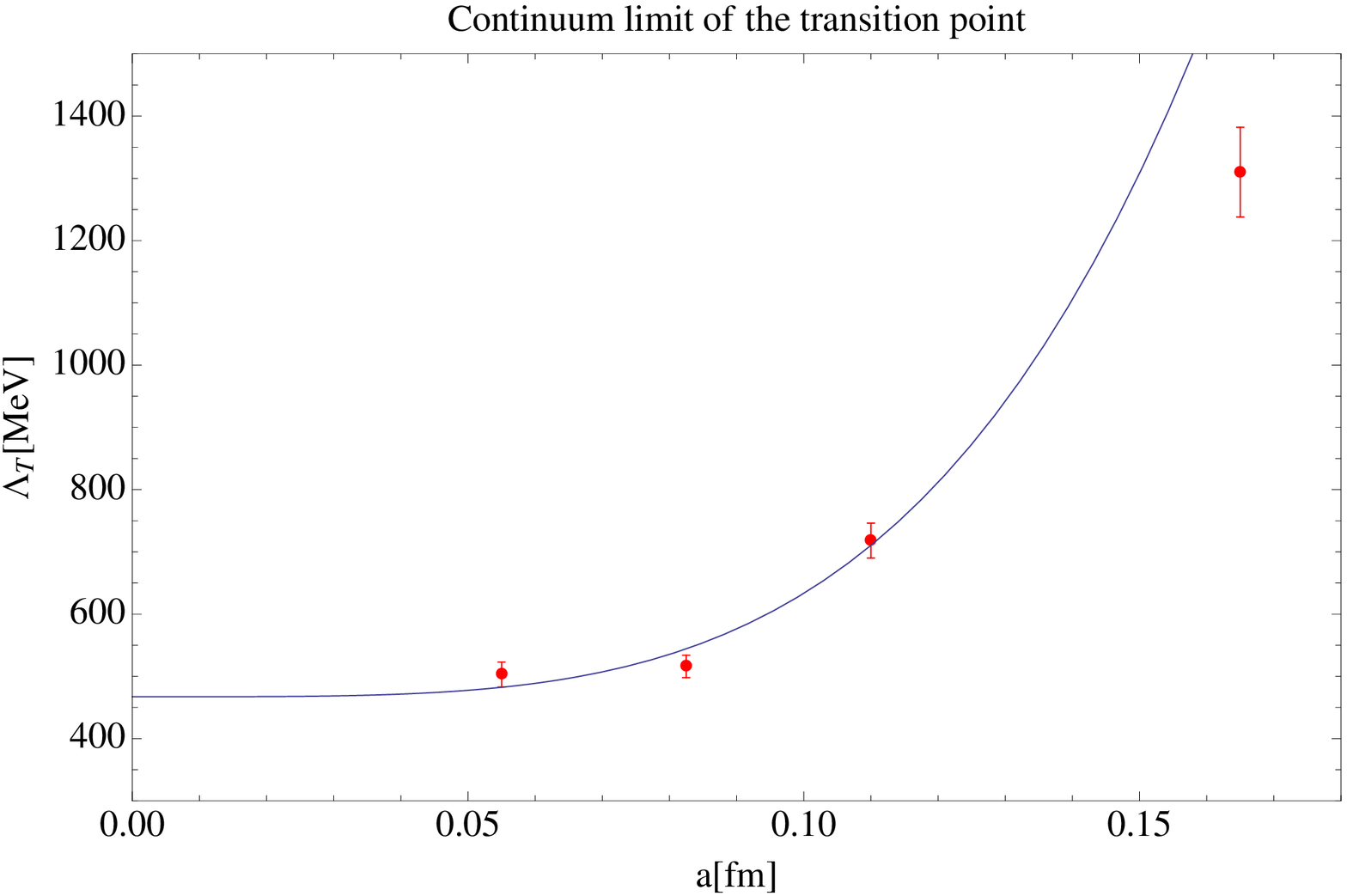}
  \vskip -0.10in
  \caption{Left panel: Determination of the chiral polarization scale,
           $\Lambda_T$, for the finest lattice.  Right panel: Dependence of the
           chiral polarization scale, $\Lambda_T(V,a)$ on the lattice spacing at
           finite volume.}
  \vskip -0.10in
  \label{C_A_and_cont}
\end{figure}

Fig.~\ref{C_A_and_cont} (left panel) shows that the dynamical correlation
coefficient decreases monotonically from positive values at low modes to
negative values at higher modes. A linear interpolation determines the scale
$\Lambda_T$, at which it crosses zero, at which point there is no dynamical
tendency for enhancement or suppression of local chirality in comparison to the
associated $X$--distribution for randomized, uncorrelated left-right
components.

We repeated the determination of this scale for each of four lattice spacings
in order to look for a continuum limit.  Fig.~\ref{C_A_and_cont} (right panel)
suggests that $\Lambda_T(V,a)$ has a finite continuum limit at fixed volume.  A
fit to the form $c_1 + c_2 a^4$, with the coarsest lattice excluded from the
fit, is drawn to guide the eye.  Further studies must address whether this
``chiral polarization scale'' survives in the infinite-volume limit, but our
check of the finite volume effect at a single lattice spacing is encouraging.

\section{Summary}

$X$--distributions are detailed probes of QCD dynamics.  The use of previous
definitions in the literature is misleading, since the distributions are
dominated by kinematical effects.  It is imperative to use the absolute
$X$--distribution as it removes kinematical effects leaving correlations
induced by QCD dynamics.  We have discovered that the absolute
$X$--distributions of the lowest several Dirac eigenmodes are convex
(polarized) and those of higher modes are concave (anti-polarized).  We have
determined the scale $\Lambda_T$ of the transition; its continuum limit is
finite (at least for finite volume) and, it is proposed, can define a "chiral
polarization" scale of QCD.  Complete details are in reference~\cite{Ale10}.
The absolute $X$--distribution was first discussed in~\cite{Dra05}.


\begin{thebibliography}{99}

\bibitem{Hor02}
  I. Horváth et al., Phys. Rev. D65 (2002) 014502.

\bibitem{Rabble}  
  T. DeGrand, A. Hasenfratz, Phys. Rev. D65 (2002) 014503; 
  R.G. Edwards, U.M. Heller, Phys. Rev. D65 (2002) 014505; 
  I. Hip et al., Phys. Rev. D65 (2002) 014506; 
  T. Blum et al., Phys. Rev. D65 (2002) 014504;
  C. Gattringer et al., Nucl. Phys. B618 (2001) 205; 
  N. Cundy, M. Teper, U. Wenger, Phys. Rev. D66 (2002) 094505; 
  C. Gattringer, Phys. Rev. Lett. 88 (2002) 22160; 
  P. Hasenfratz et al., Nucl. Phys. B643 (2002) 280.
  I. Horváth et al., Phys. Rev. D66 (2002) 034501.

\bibitem{Dra05} 
  T. Draper et al., Nucl. Phys. Proc. Suppl. (2005) 140:623-625.

\bibitem{Ale10} 
  A. Alexandru, T. Draper, I. Horváth, T. Streuer, [arXiv:1009.4451].

\end{thebibliography}
\end{document}